# Spectral Representations and Global Maps of Cellular Automata Dynamics

Theophanes E. Raptis[1]

[1] Computational Applications Group, Division of Applied Technologies, National Centre for Science and Research "Demokritos", 153 10, Aghia Paraskevi, Athens, Greece

Corresponding author: Theophanes E. Raptis (e-mail: rtheo@dat.demokritos.gr)

**Abstract**:

We present a spectral representation of any computation performed by a Cellular Automaton (CA) of arbitrary topology and dimensionality via an appropriate coding scheme in Fourier space that can be implemented in an analog machine ideally circumventing part of the overall waste heat production. We explore further consequences of this encoding and we provide a simple example based on the "Game-of-Life" where we find global maps for small lattices indicating an interesting underlying recursive structure.

Keywords:

*Cellular Automata, Game-of-Life, Convolution, Landauer Principle*

# 1. Introduction

Evolution of ALife has more than 50 years of a rich history from the first attempts of Von Neumann and Barricelli towards self-reproducing and evolving digital organisms to nowadays attempts for whole world sentient simulations. The field has recently been reviewed by Taylor in [1], [2]. The particular class of simulation tools known as Cellular Automata (CA) were first introduced by Von Neumann [3] and later popularized by Conway [4] and Wolfram [5], [6]. The later in fact represents a turn towards a rather strong conjecture in favor of the so called *Church-Turing Thesis* [7] for all of reality including perhaps biological entities both natural and synthetic as being effectively computable. Philosophical ramifications aside, it appears that for all practical purposes certain aspects of reality are well captured in various circumstances by similar discrete models which stand for good alternatives in comparison with the numerically heavy and time consuming ones based on differential equations both ordinary and partial. Well known examples include diffusion-reaction models, fire and disease spread models, lattice-gas automata and so on.

The scope of the present article is not in altering anything in the original definitions of all similar classes of automata which are self-sufficient but rather to complement their study which in many respects remains difficult due to their combinatory nature as it often happens with problems in discrete mathematics. The possibility of a complete analog implementation of both parallel and serial machines outside the ordinary Von Neumann architecture of stack-based machines will be explored in subsequent publications. In the course of developing this approach we also recover an equivalence that allows mapping a general class of discrete complex systems into a more compact form which is a subset of signal theory that does not appear in the present literature and is worth reporting independently of the other objectives. This is inherently linked to the primary objective in that a complete transcription of states into signals in a particular way, as explained in the next sections, is necessary for such a feat.

Motivation behind the main objective is based on two serious reasons of an entirely practical as much as pressing nature that are now attracting wide attention. The first is the expected breakdown of Moore's law and the need for alternative computing machinery. It is the author's opinion that this heralds also a need for reviewing analog computing but not through the old Leibnizian paradigm as that followed in the original 1945 Shannon's GPAC (General Purpose Analog Computer) [8].

The second reason is the need to reduce if not entirely eliminate the tremendous waste heat produced by the constantly expanding data center infrastructure. This is a mostly serious problem in the context of global warming and overburdening of current electrical power plants for cooling which cannot be overcome without extreme measures as the recent announcement by Google on the need to install its next such centers underwater. Even if heat recycling and other techniques were to be used, the current state of the art as well as the extreme difficulties in advancing true quantum based computing infrastructure justifies the search for alternative computing techniques that could be both less costly and more efficient. As any and all of the above reasons represent true limits to growth and thus potential evolutionary threats, the paradigm presented here could be expanded to many possible directions, equally beneficial.

The simplest model of parallel computing with synchronous automata given by the CA paradigm is here utilized to present a generic transcription method that can be realized either with existing Digital Signal Processing (DSP) or purely analog techniques. The already proven computational universality for a particular subset of even elementary CA provides sufficient motivation for their use in the particular transcription method applied here. It will be subsequently shown that any such device is effectively equivalent to a set of nonlinearly coupled filters.

In the effort to unify their study and also allow for arbitrary neighborhood topologies with higher dimensionalities in an easily understandable and more efficient way that avoids some unnecessary obfuscation of symbol tables and similar, we introduce a compact treatment which allows an analytical reformulation as a set of dynamical systems in a rather transparent way. To this aim, we present a different encoding of the dynamics as a composite of a linear and a nonlinear mapping and we concentrate on certain mathematical aspects of them that appear to be universal over a very broad class of models which may extend beyond elementary CA, independently of their particular scope.

It is equally possible to perform a similar transcription for serial machines but the process is lengthier and requires additional measures. This research is currently underway and it will be presented in a subsequent publication. Hopefully, this will also allow more advanced treatment by other researchers using the full armory of dynamical systems theory in uncovering important aspects of the dynamics in a variety of different practical situations including social simulations and artificial life studies.

In the next sections we present the necessary definitions from convolution calculus for understanding the tools developed. In section *2* we present the basic machinery behind the decomposition. In section *3* we also show an additional possibility an alternative type of dynamics appropriate for analog machines with special encoding which guarantees constant entropy during evolution of the set of all memory states. In section *4* we comment on the technical details for expanding the previous description for arbitrary dimensionality and in section *5* we use a 2D CA with a rule for the game of life to provide a toolbox for experimentation on this new technique and some results on global maps that can be easily obtained in the given framework.

## 2. CA dynamics in a dual Fourier space

While the original definition of CA is relying heavily on old nomenclature invented in the 30s by mathematicians like Turing and Post, for the general theory of Automata, the case of synchronous CA can be treated with relative ease by introducing certain tools from linear algebra and convolution calculus. We remind that ordinary CA are defined by a tuple $\{L^D, \sigma, N, T\}$ where $L$ stands for the length of a hypercubic integer lattice of dimension $D$, $N$ stands for a neighborhood of arbitrary topology around each cell on the lattice taking values from an alphabet $\sigma$ in some base $b$ arithmetic, $T$ is a transition function usually defined as a look-up table (LUT) encoding all possible combinations of cell states of any neighborhood and the associated next states of the central cell. Ordinary symmetric neighborhoods are special cases in the general set $N$ defined by a radius $r$ with a width $2r+1$ giving rise to $2^{2r+1}$ neighborhood combinations in the 1D case and similarly for higher dimensions. A set of appropriate boundary conditions is most often taken on a toroidal, multiply periodic topology which we also use here.

As a first step we recognize that any transition table can also be redefined as a 1D graph of a discrete transfer function. This is due to the fact that any symbol-wise function of many discrete variables on some bounded space is always reducible into an equivalent 1D form by taking the polynomial representation of its multi-index space as a mapping to a unique index. Thus from any such LUT of the form $T(c_1,...,c_k) \to c', \{c_n\}_{n=1}^{k}, c' \in N$ mapping neighborhood values into new cell states, one can always derive a single transfer function graph through a complete enumeration of all combinations. Given an alphabet $\sigma$ in base $b$ arithmetic any such map can always be written analytically in the form

$$T \to f : N \times \sigma \to \sigma : c' = f(c_1 + c_2 b + ... + c_{k-1} b^{k-1}) \tag{1}$$

for any and all of the $b^k$ possible rules given as symbol strings. In the above we made use of the polynomial representation for mapping all possible neighbor combinations into the set of integers in the interval $[0,...,b^k - 1]$.

Before analyzing the role of higher dimensionality for the lattice itself, we restrict attention to the case of a 1D CA. It is then possible to utilize a particular addressing scheme for (1) which leads directly to a decomposition of the dynamics in two separate maps as $L_{n+1} = (f \circ g)(L_n)$ where $f$ stands for the generally nonlinear part and $g$ is a strictly linear map. We do this by introducing a generic *Interaction Kernel* as a special $L \times L$ circulant matrix of the form

$$\mathbf{C} = \begin{bmatrix} s_1 b & s_2 b^2 & \cdots & s_0 \\ s_0 & s_1 b & \cdots & s_{L-1} b^{L-1} \\ \vdots & & \ddots & \vdots \\ s_{L-1} b^{L-1} & \cdots & s_2 b & s_1 \end{bmatrix} \tag{2}$$

Any such matrix is composed from the cyclic permutations of the vector $\mathbf{c} = [s_0, s_1 b, ..., s_{L-1} b^{L-1}]$ where we use the binary symbols $s_i$ to denote the topology of any possible neighborhood via a logical mask plus a wraparound at the matrix edges for the cyclic boundary conditions. Any such matrix is a special case of a Toeplitz matrix and it is fairly easy to realize it from a single vector in the first row henceforth referred as the "kernel vector" wherever a generic routine for constructing Toeplitz matrices is available. In Matlab, the unique command *toeplitz*( *c*(1), fliplr(*c*(2:end)), *c* ) is sufficient yet, even this is not necessary. As we will show, a particularly elegant solution is offered by the general diagonal form from which one deduces that only the DFT over the defining first row is actually necessary and this technique was used in some of the example codes. This is a ubiquitous technique in modern signal processing and image analysis and it is further discussed in the case of higher dimensional automata in section 4.

We notice the existence of two important limits which we will term for convenience the "*Holographic*" and "*Anti-holographic*" limits with the first corresponding to the case of the binary mask digits all ones (global coupling) and the second to the minimal case of only $1^{st}$ order nearest neighbors. The binary mask allows even for random, disconnected ("non-local") neighborhoods. In such a case the *C* matrix will obtain a sparse band structure. In the most trivial case of nearest neighbors apparently one only has a sparse matrix of a single 3 entries band around the main diagonal..

The above definitions allow decomposing the dynamics of any 1D CA as a set of two discrete evolution equations in the form

$$h_n = \mathbf{C} \cdot L_n$$
$$L_{n+1} = f(h_n)$$
(3)

We now note that the above picture coincides with that of a special class of filters with a circulant connectivity matrix of weights given by the kernel *C* and *h* playing the role of what is known in neural network terminology as the "excitation field". The new variable *h* stands for an intermediate addressing field which is passed through the transfer function with the method given by (1) naturally coded in the interaction kernel. In practice, one does not have to actually store a large array in memory due to the internal symmetry of the kernel. Thus the operation defined by the first evolution equation in (3) can be realized as a loop of dot products of two vectors followed by a one-step cyclic shift to the right spending no more than $2L$ memory positions. To gain a deeper understanding of the kind of dynamics, we have to analyze the special properties of such circulant kernels.

There are very many special properties of circulant matrices [9], [10] which are expressed analytically as integral kernels of a convolution operator on cyclic groups and they are often used in cryptographic applications. First we notice the property of circular convolution for any circulant kernel which allows writing $\tilde{h}_n = \tilde{\mathbf{c}} * \tilde{L}_n$ where one simply takes the component-wise algebraic product denoted by * of the separate discrete Fourier transforms (DFT) of both the vector *c* defining the original kernel as well as the lattice state vector in their cyclic extensions so as to match lengths. We also notice that this is associated with the adjoint operation of cross-correlation which for finite sequences is just a conjugation of $c$ given by $\tilde{\mathbf{c}} * \otimes \tilde{L}_n$. For finite, real symbols the two are intimately related by a mirror inversion. Various known recursive methods exist for such operations [11] in both analog and digital electronics.

It is possible to express the action of the interaction kernel as a sequence of cyclic permutations in the form

$$h_n = \sum_{i=1}^{L} s_i b^i \overline{\mathbf{X}}_n^{(i)}$$
$$\overline{\mathbf{X}}_n^{(i)} = \pi_0^i \cdot L_n$$
(4)

In the above, we used the irreducible representation of the cyclic permutation via the matrix

$$\pi_0 = \begin{bmatrix} 0 & 0 \cdots & 1 \\ 1 & 0 \ldots & 0 \\ 0 & \ddots .1 & 0 \end{bmatrix}$$

The superposition of different copies in (4) is just shorthand of what a programmer actually does by keeping separate copies of the lattice in memory in order to compute a neighborhood's status. We further explore some mathematical properties of the decomposition in (3) that are relevant for possible DSP or analog based implementations.

We are in fact in a position to further compact the first operation in (3) using the second important property of circulant matrices according to which all such matrices are diagonalized by a discrete Fourier (DFT) matrix as $\mathbf{C} = \mathbf{F}^{-1} \lambda_C \mathbf{F}$. The eigenvalue matrices are always given in terms of the defining vector $c$ and powers of the $L$ roots of unity as

$$\lambda_C = \sum_{i=1}^{L} s_i b^i \hat{\Omega}^i$$
$$\hat{\Omega} = \begin{bmatrix} \omega_1 & 0 \cdots & 0 \\ 0 & \ddots & 0 \\ 0 & 0 \cdots & \omega_L \end{bmatrix}$$
(5)

The above unique property allows rewriting the dynamics in (3) in its dual form

$$\tilde{h}_n = \lambda_C \cdot \tilde{L}_n = \sum_{i=1}^{L} s_i b^i \overline{\mathbf{L}}_n^i$$

$$L_{n+1} = f(h_n), \quad \overline{\mathbf{L}}_n^i = \hat{\Omega}^i \tilde{L}_n \tag{6}$$

In (6) we have taken the discrete transforms of both the addressing field and the lattice as $\tilde{h} = \mathbf{F} \cdot h, \tilde{L} = \mathbf{F} \cdot L$ at any one time. This also leads to the interpretation of the total dynamics as a sequence of homomorphisms

$$\tilde{L}_{n+1} = \left(\hat{\mathbf{F}} \circ \hat{f} \circ \hat{\mathbf{F}}^{-1}\right)(\lambda_C \cdot \tilde{L}_n) \tag{7}$$

We may as well use the inverted picture

$$\tilde{h}_{n+1} = \lambda_C \cdot \left(\hat{\mathbf{F}} \circ \hat{f} \circ \hat{\mathbf{F}}^{-1}\right)(\tilde{h}_n) \tag{8}$$

Using (5) in (8) allows extracting the superposition

$$\tilde{h}_{n+1} = \sum_{i=1}^{L} (s_i b^i \hat{\mathbf{M}}_i)(\tilde{h}_n)$$

$$\hat{\mathbf{M}}_i = \hat{\Omega}^i \left(\hat{\mathbf{F}} \circ \hat{f} \circ \hat{\mathbf{F}}^{-1}\right) \tag{9}$$

The dynamics for the dual picture of the original lattice in (7) can be decomposed in terms of the scaled images given by $\overline{\mathbf{L}}_n^i$ but due to the highly nonlinear nature of the $f$ map the functional operator cannot be passed inside the summand. Interpretation of the $\overline{\mathbf{L}}_n^i$ terms can be given in terms of dense samplings of the unit circle by the $\Omega$ matrix and its powers. This is then equivalent to a harmonic function of wavelength $L$ which modulates the Fourier components with its powers applying a phase shift. This behaves in a similar manner to the case of a transmission line with a single travelling harmonic wave filtered by the neighborhood which acts as a window function. In the holographic limit of the whole lattice neighborhood, the phase shifts from left to right scan the lattice as a travelling wave. There appears to be a formal similarity of the dual picture with a sampling of a discrete set of signals.

In the final picture we see that the same dynamics can be represented as a unique superposition of signals under the action of certain operators on the original signal. In order to be able to utilize this observation for implementing an analog machine without switching and memory erasure operations responsible for waste heat production, a special technique will have to be introduced in the next section where the intermediate $f$ function which here stands only for a look-up table (LUT) will be turned into a new linear matrix operator of which the elements are modulated by the circulant filter. The essential nonlinearity is then transferred into the modulation scheme as an exponentiation operation.

3. **The "Wheel" implementation via permutations**

We first introduce special constant entropy encoding which can be of wider interest. Consider an arbitrary message in some alphabet in base *b* composed of words $w_i$ of length *n*. Each word is characterized by its own Shannon entropy $s(w_i)$. It is then possible to construct a new code of *bn* letters henceforth referred to as the "wheel encoding", such that the total entropy per word remains constant. The proof can be given constructively. Assume a repetition of the original alphabet as a set of *n* sorted lists of the form $l = [0,...,b-1]$. For any word $w_i = \sigma_1,...,\sigma_n$ we then associate each letter $\sigma_i$ with a cyclic permutation of exactly $\sigma_i$ steps for each of the *n* sorted lists of *L*. The final list contain exactly the same message with an entropy which remains constant given that the number of symbols remains always the same with a value of $\log b$. We then proceed by applying the wheel encoding to the original CA lattice.

Indeed, by noticing that the sole action of *f* in (3) as well as in (8) and (9) is to pick up a particular symbol from an ordered list of symbols we can replace it with a special matrix vector operation. We first introduce a constant vector $\mathbf{r} = [0,1,...,b-1]$ as a list of all the symbols from a b-ary alphabet and its L-periodic extension $\mathbf{R} = \{\mathbf{r}\}^L$. We then construct a *Lb* x *Lb* matrix of *b* x *b* permutation blocks as

$$\mathbf{K} = \begin{bmatrix} P_1 & 0 & \cdots & 0 \\ 0 & P_2 & 0 & \vdots \\ \vdots & 0 & \ddots & 0 \\ 0 & \cdots & 0 & P_L \end{bmatrix} \quad (10)$$

All permutation blocks are now given as

$$P_i = \pi_0^{f(h_n(i))}, i = 1,\ldots,L \quad (11)$$

Next we reformulate the original kernel vector so as to keep track of only the relevant positions, for instance by adding *b-1* zeros in the form $\mathbf{c} = [s_0,0,...,0,s_1 b,0,...,s_L b^L,0,...,0]$. Thus effectively the same length *L* CA is realized. Then, the overall operation takes the form

$$L_{n+1} = \hat{\mathbf{K}}(f(h_n)) \cdot \mathbf{R} = \hat{\mathbf{K}}\left(f\left(\hat{\mathbf{C}} \cdot L_n\right)\right) \cdot \mathbf{R} \quad (12)$$

We notice that the type of nonlinearity induced by (11) actually represents a hyper-operation and in particular a hyper-exponentiation [12] between maps from a vector space to another. In the simplest of cases, one calls a sequence of such operations tetrations, or Power Towers. When applied on the same complex number they have an infinite limit given in terms of the Lambert W function [13]. Some other generalizations have recently been studied by Galidakis [14], [15]. The case of a hyper-exponentiation in a sequence of mappings seems to be much less explored but this lies outside of the scope of the present article.

A more straightforward picture in the inverted address space can be given as

$$h_{n+1} = \hat{\mathbf{C}} \cdot \hat{\mathbf{K}}_{f(h_n)} \cdot \mathbf{R} = \hat{\mathbf{F}}^{-1} \lambda_C \mathbf{F} \cdot (\hat{\mathbf{K}}_{f(h_n)} \cdot \mathbf{R}) \qquad (13)$$

Given any CA that is considered useful for a particular task and perhaps universal computation, one can always encode the initial condition in (13) by a single application of the corresponding circular kernel and obtain the desired result at a certain step *n* assuming a method exists to define a halting state directly in the address space.

The picture of the extended dynamics offered by (12) and (13) is that of a set of wheels of prescribed symbols in an extended tape of length *Lb* out of which only *L* symbols are "observed". This is in fact equivalent to a set of *L* wheels just as in the case of one-arm-bandit machines with a slot. The role of the observation slot is here played by the extended kernel defined by the *c* vector. We can bring the dynamics of (13) to its equivalent dual form in the Fourier domain

$$\tilde{h}_{n+1} = \lambda_C \mathbf{F} \cdot \left( \hat{\mathbf{K}}(g_n) \cdot \mathbf{R} \right) \qquad (14)$$

In (14) we used the shorthand $g = f(h) = f\left(\mathbf{F}^{-1} \cdot \tilde{h}\right)$. From the expression of the eigenvalues in (5) we also have the equivalent

$$\tilde{h}_{n+1} = \sum_{i=1}^{L} s_i b^i \left( \hat{\Omega}^i \cdot \hat{\mathbf{F}} \right) \cdot \left( \hat{\mathbf{K}}(g_n) \cdot \mathbf{R} \right) \qquad (15)$$

As a final result, (14) or (15) describes the reconstruction of a signal from permutations of a constant vector or "*program signal*" controlled from its previous image. In this new encoding, initial conditions are fed through a particular choice of exponents for the cyclic shifts in *K*.

Regarding possible advantages of the circular implementation of the dynamics, we notice that the "one-arm-bandit" trick is essentially identical with a set of *L* wires carrying harmonic electrical signals of the same frequency with different relative phases. Taking a *b*-sampling of the phase unit circle we get to an equivalence of the original alphabet with *b* complex roots

of unity. Indeed, for a number of signals $s_i(t)$ the original dynamics could be written in the form

$$\mathbf{s}_i(t+\delta t) = \exp\left[\frac{2\pi \mathbf{i}}{b} f\left(\mathbf{C}_{ji} \cdot \text{Im}(Log\mathbf{s}(t))_i\right)\right] \qquad (16)$$

Effectively, the action of the original kernel corresponds to a multiplexing scheme and the action of the transfer function becomes equivalent to a phase-shifting device. Although faithful, the representation used in (16) is rather difficult technically due to the nonlinear phase extraction operation. Instead, we can employ a much simpler method avoiding such nonlinearities via a different type of encoding.

Assume then, a single signal where instead of the phase sampling, one utilizes a periodic repetition of the same frequency comb characterized by an equipartition of the frequency domain over an interval $[\omega_0,...,\omega_{b-1}]$ with successive spectral amplitudes forming a linearly increasing staircase. In such a configuration, rotation of the "wheels" or the equivalent permutation blocks in (10) and (11) becomes identical with a rotation of the amplitudes at these frequencies in each sub-interval. In fact, single nonzero amplitude at a specific position associated with a symbol would suffice. Using an FFT filter and a LUT one can realize any change of such an ideal signal as $\mathbf{s}(t+\delta t) = \mathbf{s}(t) - \exp(\mathbf{i}\omega_i t) + \exp(\mathbf{i}\omega_j t)$ for any $i \rightarrow j$ transition. This is a trivial operation that does not alter the total spectral density, a fact that allows one to speak of isospectral and subsequently constant entropy computation at least in its ideal form.

Part of the waste heat production in existing digital devices occurs as a result of the well known "Landauer's Principle" [16], [17] as a consequence of the 2[nd] thermodynamic law and according to which, any irreversible elementary digital operation like a bit erasure has a lower bound of entropy increase of the order of $K \log 2$ with K being the Boltzmann constant. The true reason that the construct presented here manages to circumvent this is a consequence of general principles in by now well established Reversible Computing as discussed by Perumalla [18], especially in Chapter II-5 regarding the principle of no memory increase. Additionally, as also mentioned there one understands that the equivalent of the entropy increase must have been transferred in some other entropic function. Indeed, in our case this is now accumulated in the computation history paths and in particular, to the "which-slot" question with respect to the preimages of which ignorance increases due to the original non-invertibility. Hence, nothing is violated from a purely thermodynamic point of view.

Engineering optimization concerns among others, ohmic losses due to extensive use of wiring, and the introduction of alternative analog implementations of the type presented here have certain appealing characteristics including possible speed improvements and probably, also, lower total heat production due also to the minimization in terms of single signal ideally and/or lesser wiring required where the term wire here may also mean a specially designed set of transmission lines. Given also that many CA have been shown to exhibit computational universality, like the 2D case of the famous "Game of Life", it could be important to know how to encode their dynamics in the extended 1D signal form presented here.

4. **Significance of dimensional reduction for the composite map**

The full decomposition of the dynamics for 1D CA can be utilized to achieve an ultimate dimensional reduction for any $L^D$ lattice via a special expansion of the interaction kernel. The starting point is based on the observation that as long as the interaction kernel keeps track of all possible correlations represented by a neighborhood of arbitrary topology the same can be done with neighborhoods that are spread across various directions in a higher dimensional space thus making this representation insensitive to the particular dimensionality of the original lattice.

To give an example, the easiest way to hardwire all the relevant topological information in a kernel can be given with a 2D $L \times L$ cell lattice. One has to reshape the original matrix as an $L^2$ vector reading the lattice in a row-wise fashion. Afterwards, the kernel can be set up with the aid of an appropriately defined vector $c$ of length $L^2$ such that the binary mask $s$ will have ones corresponding to only those positions that match the original 2D neighborhood thus preserving the original correlations. In order to preserve the full toroidal topology, the defining vector $c$ can be setup with a double periodicity as follows.

Given an original sub-vector of length $L$ with code $\mathbf{c}_0 = [s_1 b, s_2 b^2, ..., s_0]$ one takes a $L^2$ extension $\mathbf{c} = [b^k \mathbf{c}_0, b^{2k} \mathbf{c}_0, 0, ..., 0, \mathbf{c}_0]$. The key point is that one now has a double rotation corresponding to the two row-wise and column-wise periodicities which in the 1D unfolding manifest themselves as a block circulant structure in the final kernel where each $L \times L$ sub-matrix is circulant but the blocks also rotate. To construct such a matrix we can apply a recursive procedure where we first construct a circulant of symbols for each submatrix. Then we can replace each symbol with another circulant block. This procedure can be summarized with the use of the Kronecker tensor product between an initial circulant block $c_0$ and a new one $c_i$ for each dimension containing the "gaps" in powers of 2 from the lattice dimensional reduction. In the case of a 2D CA one simply takes the product $\mathbf{c}_1 \otimes \mathbf{c}_0$ where

$$\mathbf{c}_0 = \begin{pmatrix} 2 & 4 & 0...0 & 1 \\ 1 & 2 & 4 & 0...0 \\ 0 & \ddots & \ddots & \ddots 4 \\ 4 & \ddots 0 & 1 & 2 \end{pmatrix}, \quad \mathbf{c}_1 = \begin{pmatrix} 32 & 64 & 0...0 & 1 \\ 1 & 32 & 64 & 0...0 \\ 0 & \ddots & \ddots & \ddots 64 \\ 64 & \ddots 0 & 1 & 32 \end{pmatrix}$$

Specifically, the whole procedure if repeated for n dimensions always results in a special class of bandpass filters with an arithmetic fractal structure as shown in the next section with example code.

The above defined class generalizes a well known practice in image processing and especially in image deblurring as in ref. [ 19 ] where various coding examples are also offered, as well as in the theory of random fields [20]. The general class of matrices for various boundary conditions is there known as the Block Circulant with Circulant Blocks (BCCB) and other combinations like Toeplitz sub-blocks (BTTB) which are also reducible into the large BCCB ensemble. By some general theorems in linear algebra [21] we know that all these classes of matrices of order $L^D$ x $L^D$ can be reduced to taking the tensor product of DFT matrices of the associated block dimensions over the input vector and its conjugate inverse after multiplying with the eigenvalues. The eigenvalues vector is always precomputed from the total DFT of the first column of all circulant blocks. This results in a runtime complexity which for square lattices with a power of 2 side length should scale as $O(2^{2k+1}k)$ floating point operations.

The above observations are not just a mere technical convenience but they also allow a physical reinterpretation of the 2-step decomposition shown here. As a matter of fact, in the general theory of deblurring, the kernel vector introduced here is interpreted as a discretization of the optical Point Spread Function (PSF). The new generalized class presented here corresponds to the case of inhomogeneous blurring from a fractal material. As a result, our choice of decomposing the dynamics leads to interpreting $f$ as an illumination function from which the image of the automaton is mapped to an intensity function before being filtered by the rule function.

A toy model could be experimentally constructed using the recently introduced MIT morphing table by Net Media Lab in the inFORM project [22] using appropriate light sources as a kind of Video Feedback system. For the implementation of an analog machine though, it is more interesting to transfer the same technique into the frequency domain where Fourier components allow a mostly compact representation by mapping the whole cell space on a single composite signal or at least bunches of signals with appropriate multiplexing techniques. Similarly, a wavelength based device could be envisioned as a special cavity with time-dependent controllable boundary conditions for experimentation or even an attempt to directly transfer the encoding of section 3 to the reciprocal lattice (momentum space) of an appropriate solid state material. This generic model stands for a kind of a "*Pluribus Unum*" principle for similar discrete models of complexity which may prove of wider significance in their study by analytical methods too.

**5. Computational experimentation with the "Game-of-Life".**

All the examples in this section are currently available from the author's github account [28]. Simple applications for 1D CA with "*Spectral1D.m*" and "*Wheels1D.m*" show the compactness and minimal complexity of this coding technique. Extensive use of vectorization allows minimal code complexity with a one step evaluation of whole arrays. Moreover, many other less elementary CA classes could be explored with minimal effort and alterations. This includes the more difficult case of 3D CA including Ising models where visualization and inspection of results as well as isolation of interesting structures become tedious. The purpose of this section is to explain the methods for this with some simple known examples and not to perform such a full exploration which would have to go to great lengths given also the non-polynomial (NP) nature of computation for the state space of all possible configurations.

In the case of 2D CA it is necessary to translate any rule to its unfolded form as a 1D numerical array. To this aim, one either has to take a lingual description and translate it in a functional form or as in the case of the game-of-life CA to turn this in to a form of transition table in ternary coding where 0 states are states of no alteration while ±1 states are just added to 0 or 1 states to change them into their complements. This is achieved by taking the *xor* of the central cell and its next state from the original rule to isolate actual transitions and changing encoding to a {-1, 0, 1} alphabet.

To complete the transcription of the Life CA rule, we isolate the central cell value so as to be the leading bit in the polynomial representation of the neighborhood state. This representation utilizes the special property that for the whole interval of $[0,...,2^9]$ states, the first half interval up to 255 represents "dead" central cells and the other half up to 511 "alive" ones. The particular choice affects also the kernel structure is a way explained in the next paragraph. Completing the unfolding is possible with a special function over the integers known as the Digit-Sum function [23]. This was used to provide the number of living cells as the number of one bits. Given the function $S_2(n), n \in N$ the Life CA rule can be written analytically as a Boolean function of the form

$$r(n) = \begin{cases} S_2(n) = 3, & n \in [0, 255] \\ (S_2(n) = 3) \wedge (S_2(n) = 4), n \in [256, 511] \end{cases} \quad (17)$$

The $S_2$ function is easier to compute independently over a succession of exponential intervals where a single new bit is added using a recursion as

$$S_D(0) = \{0\}, S_D(0...2^{k+1}) \leftarrow \{S_D(0...2^k), S_D(0...2^k) + 1\} \quad (18)$$

Using this method, one obtains the results of figure 1 where live output cells are shown with a bar graph. In the liferule code we also computed the transition map in ternary code. Explicit form is given again analytically as

$$T_r(n) = \begin{cases} r(n), & n \in [0, 255] \\ 2(r(n) \oplus 1) - 1, n \in [256, 511] \end{cases} \quad (19)$$

The sole difference between the two in practice is that the first computes the next central cell values as $c'_i \leftarrow r(h_i(c_i))$ where $h_i$ the address field, while the transition map is additive as $c'_i \leftarrow c_i + T(h_i(c_i))$. The second allows extracting a discrete time derivative for the evolution. This type of transfer function shows an alternative analog implementation as it has the form of a discretized integro-differential operator. The simplest possible implementation of a synthi machine could then be given with the aid of a second stage responsible for performing additive synthesis. This can be described analytically as a diagonal matrix $\mathbf{K}$ with $diag(\mathbf{K}) = T(h)$ and an oscillator bank providing a constant signal array $s_0$ with the total signal given as a dynamical system of the form

$$s(t + \tau) = s(t) + \mathbf{1}^\mathrm{T} \cdot \mathbf{K}(h) \cdot s_0(t) \tag{20}$$

The same scheme can also be applied for a serial machine with only a pair of frequencies been affected each time. In the case of a wavelength encoding, (20) would correspond to appropriate mode clipping.

We also provide two examples with an explicit construction of the kernel matrix which is given by the "*kernel.m*" routine as well as the more general "*kernelND.m*" which utilizes the Kronecker product for the creation of equivalent kernels in arbitrary dimensions. These can be run independently given the one side length of a square lattice with a particular encoding suited for extracting the transition map and study the eigenvalue and eigenvector spectrum in various dimensions. It helps visualizing the fractal structure due to the use of the Kronecker product. Both routines "*liferule.m*" and "*kernel.m*" are called at the initialization stage. An additional difficulty here stems from the use of a permutation of the central cell position in the rule encoding introduced previously. As this affects only the central diagonal blocks this is taken care with a sum of Kronecker products. Given two matrix blocks $c_1$ and $c_2$ the kernel is now constructed as $c_1 \otimes c_0 + \mathbf{I} \otimes c_2$ where $\mathbf{I}$ stands for the $L \times L$ identity matrix. This is the sole reason for not using directly the eigenvalues for evolving the system as in the 1D case as it results in two different Kronecker products of DFT matric. Internal structure of a kernel's sparse matrix is shown in figure 2. The choice of the rule encoding is of course entirely up to the user and one may use the original kernel form without the permutation.

The main code "*liveon.m*" contains the simplest possible realization of a 1D encoding for a 2D lattice. This can always be returned into a square matrix shape via the MATLAB command "*reshape*". The structure of the kernel is what guarantees that no topological information or correlations during evolution get lost and three explicit examples with two known periodic oscillators of periods 2 and 5 respectively, as well as a long term evolution "*Methuselah*" pattern known as "acorn" is provided as predefined choices. That all information must be preserved is of course already assured given that proofs in sections 2 and 3 provide all necessary and sufficient conditions before the choice of implementation.

Additionally, we provide example code for retrieving global maps of the dynamics which may be of substantial value in the overall study of the statistical properties of general CA classes. Their significance stems from the fact that they represent generalized LUTs or graphs of a single fan-in/fan-out computation for the configuration of an entire lattice state encoded into single large integers assuming infinite precision arithmetic. Having obtained these maps once allows running any arbitrary experiments for any initial condition as in the case of single number sequence just like in the case of discrete maps like the logistic, modeling a 1D complex system. This speed-up of course requires a trade-off with memory expenditure for pre-storing large chunks of state space whole lattice transitions.

Reading of the contained bit structure in large integer representation of lattices in the binary case can be sped up with the use of a Division-free Binary Decoder of which a functional version is already used inside the main codes for rule bit extraction. To facilitate the reader, a fast version using solely bitwise operations for the combined binary logarithm and a power-of-2 detector is separately provided in the DFFBD module. The basic principle behind it is the minimal perturbation of an integer's bit structure when adding a single bit (binary shift) at a zero position or an isolated bit in which case $S_2$ changes either by 1 or remains invariant respectively. A *xor* operation between shifted positions by one forth and back is then utilized as a Boolean Laplacian or "edge detector" to discern changes in the bit structure.

To construct an exhaustive global map for a 2D CA square lattice of side length $L$ we consider the ordered set of strings over an interval $[0,\ldots,2^L]$ as an $L^2 \times 2^{L^2}$ array **L**. Application of the convolution operator is possible in principle directly upon all elements of *L* so as to be able to check the properties of the address field $h$ independently of any particular rule realized by the transfer function *f*. To check this we construct the equivalent map given as

$$\mu : \sigma^N \times \sigma \to \sigma' \supset \sigma : \mu(|L_n|) = \mathbf{C} \cdot \mathbf{L} \tag{21}$$

In (20) $|L_n|, n \in [0,\ldots,b^L]$ stands for the equivalent integer encoding of each particular configuration of cell states via the polynomial representation. Given a previous knowledge of (19), realization of any automaton would reduce in the study of the particular eigenvectors and eigenvalues of the permutation blocks of (11) given as

$$P_i = \pi_0^{f\left(\mu_n^i(|L_n|)\right)} \tag{22}$$

The code "*globalmap.m*" computes (21) for the binary case. Due to the NP character of this problem, computation may severely slow down after small lengths with $L = 3$ or 4. Results for the set of all $2^9$ and $2^{16}$ possible lattice states obtained this way are shown in figures 2a and 2b. We notice the appearance of an arithmetic fractal structure with increasing lattice site. This is not unique or pertinent to the particular rule of the Life game as one may see by playing with different variations and alterations in the bits from the *liferule* output, but the full identification and classification of all such possible structures even for the set of all $2^{510}$ non-trivial elementary CA rules is clearly an NP problem which calls for super-computing infrastructure, yet flexibility of existing code may allow this to be done with special programming techniques including use of GPU/CUDA facilities in dedicated machine clusters. It is possible that the set of all such global maps hides an as yet unknown global recursive structure.

**6. Discussion and conclusions**

In the present report, we offered a unifying decomposition of CA dynamics which can be used in a variety of settings with arbitrary parameterization including dimensionality, neighborhood structure and the basis of the alphabet for the cell states. We also offered an equivalent implementation which can be put into a direct correspondence with a set of electrical signals suggesting the existence of ideal analog machines carrying out the equivalent of digital/symbolic computations in a manner that allows for constant entropy and spectrum computations in an effort to circumventing waste heat production associated with both Landauer's principle for irreversible digital computation as well as the amount of wiring required.

While this procedure may appear contrived, it is often necessary to generalize to arbitrary or even stochastic neighborhoods that can also change with time if it is to take full advantage of the power of CA beyond mere elementary automata for modeling in social simulations and other cases of multi-agent systems. In such a case there is no particular advantage in keeping the standard elementary topology. Any arbitrary neighborhood can be translated into a logical mask once at the beginning of computation with no additional cost. This scheme is easily adaptable even in the case of time dependent neighborhood topology. This also allows greater flexibility in case one would be interested in applying genetic algorithm techniques for optimizing neighborhoods in particular applications with emphasis on the minimal coding complexity and need for reparametrization and recycling of existing code.

From the point of view of signal theory, given the isomorphism proven in sections 3 and 4, one always has an equivalent picture of a single signal upon which dimensionality is enforced or imposed by the interaction kernel. Hence, the message of this dual picture is that dimensionality for discrete systems can always be interpreted in two ways with the latter assigning dimensionality in the axioms defining the interactions rather than the elements themselves. This way it also introduces a new paradigmatic framework absolutely equivalent to the original but with the advantage of bringing in the additional toolbox of harmonic analysis opening two parallel ways of research one of practical benefit and the other of a more theoretical nature. Last but not least, the author would like to point out the interesting analogy between the "which-slot" question of section 3 and the "which-way" question of standard double slit interferometry as well as the analogy of the dual description with that of Heisenberg and Schrödinger pictures in quantum dynamics. We notice that analogies with quantum systems were also reported earlier by Dresden in the study of the well known Kac-ring model [24] and recently by Elze [25] in the context of so called emergent quantum mechanics [26].

Given that the simple examples of the Game of Life have already been proven to be Turing complete as well as certain generalizations known as "Collision-based Computing" [27] the author suggests that the subtle relation between wavy structures and certain types of signaling often called "gliders" and their initial conditions should be subsequently studied in detail in relation with the properties of their analog implementation including representations of harmonic and permutation groups which points towards a rich and interesting algebraic structure.

**Fig. 1:** Example of a Boolean transfer function from an unfolded transition rule for the "Life" CA.

**Fig. 2:** Structure of the interaction kernel for the 2D case of a 100 x 100 lattice.

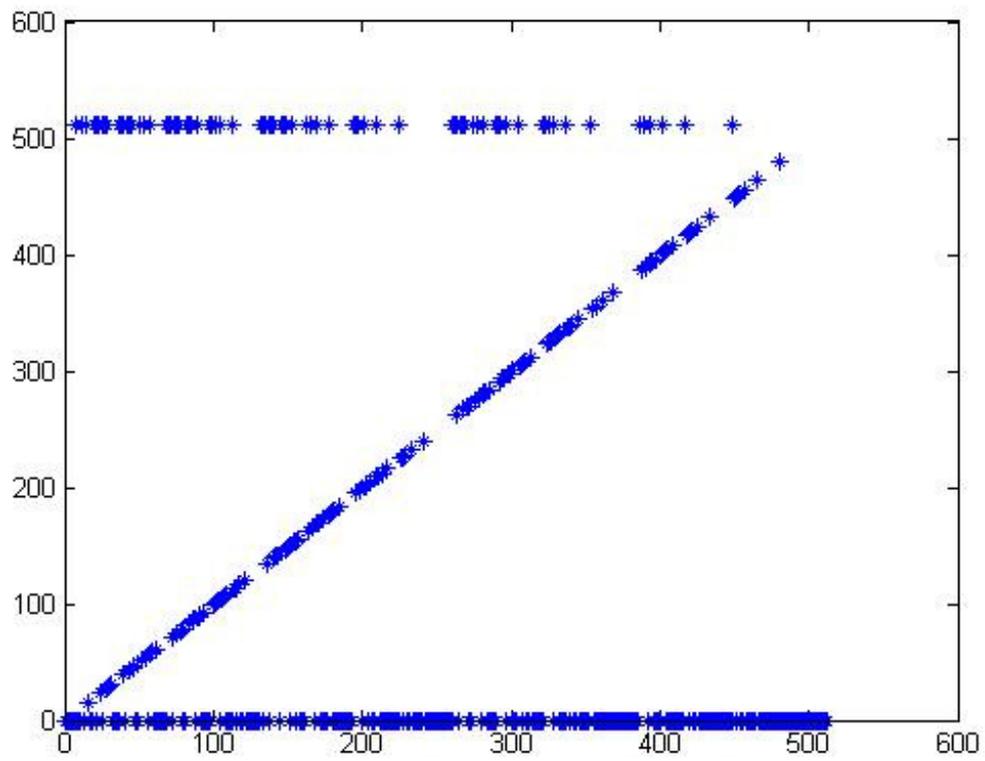

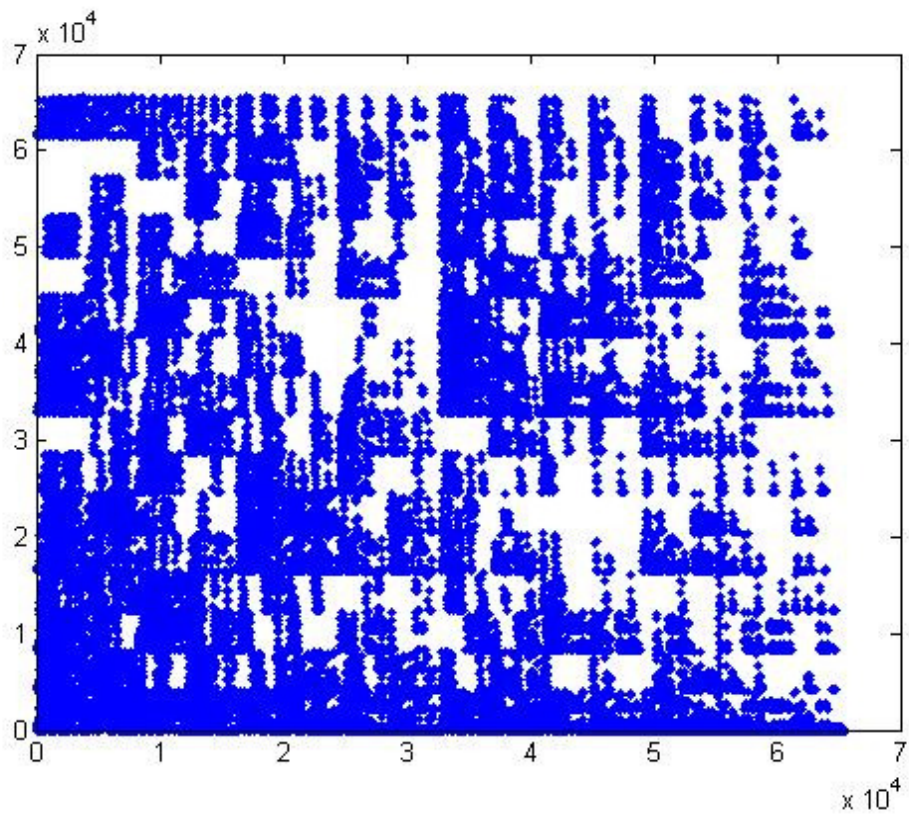

**Fig. 3:** (a) Global map of all input strings for a 9 cell (3 x 3) 2d lattice (b) the same for a 4 x 4 lattice.